# SEVEN KEY DRIVERS TO ONLINE RETAILING GROWTH IN KSA

Rayed AlGhamdi and Steve Drew
*School of Information and Communication Technology*
*Griffith University, Brisbane, Australia*

**ABSTRACT**

Retailers in Saudi Arabia have been reserved in their adoption of electronically delivered aspects of their business. This paper reports research that identifies and explores key issues to enhance the diffusion of online retailing in Saudi Arabia. Despite the fact that Saudi Arabia has the largest and fastest growth of ICT marketplaces in the Arab region, e-commerce activities are not progressing at the same speed. Only very few Saudi companies, mostly medium and large companies from the manufacturing sector, are involved in e-commerce implementation. Based on qualitative data collected by conducting interviews with 16 retailers and 16 potential customers in Saudi Arabia, 7 key drivers to online retailing diffusion in Saudi Arabia are identified. These key drivers are government support, providing trustworthy and secure online payments options, provision of individual house mailboxes, providing high speed Internet connection at low cost, providing educational programs, the success of bricks-and-clicks model, and competitive prices.

## 1. INTRODUCTION

For Saudi Arabia, e-commerce is a new wave in the information technology revolution. Since the 1990s many businesses around the world have introduced online business models as part of their operations seeking the many advantages that the online marketplace can provide (Laudon and Traver 2007). Saudi Arabia as the largest world oil producer (CIA 2009) is now seeking to develop strong information based economic structure part of which is to realise the advantages of e-commerce tools and applications (Information Centre - Saudi Ministry of Commerce 2006). Despite the fact that Saudi Arabia has the largest and fastest growth of ICT marketplaces in the Arab region (Saudi Ministry of Commerce 2001, Alotaibi and Alzahrani 2003, U.S. Commercial Services 2008 and Alfuraih 2008), e-commerce activities are not progressing at the same speed (Al-Otaibi and Al-Zahrani 2003, Albadr 2003, Aladwani 2003, CITC, 2007, and Agamdi 2008). Comparing electronic commerce in Saudi Arabia to the global volume, it is low at $150 million (U.S. Commercial Services 2008). By the end of 2007, the number of officially registered Saudi commercial organizations is 695,200; however, Saudi Arabia is still a late adopter in e-commerce field (Agamdi, 2008). Only 9% of Saudi commercial organizations, mostly medium and large companies from the manufacturing sector, are "involved in e-commerce implementation" (CITC, 2007). The question that should be asked here is "How to accelerate the adoption of online retailing in Saudi Arabia"?

## 2. REVIEW OF E-COMMERCE IN KSA

In 2001, the Saudi Ministry of Commerce established a permanent technical committee for e-commerce including members from the Ministries of Commerce, Communication and Information Technology and Finance. It also includes members from the Saudi Arabian Monetary Authority (SAMA) and King Abdulaziz City for Science and Technology (KACST) (Saudi Ministry of Commerce 2001). The roles of this committee are to follow the developments in the field of e-commerce and take the necessary steps to keep pace with them. The committee will learn from international experiences in this area, identifying the particular needs and requirements to take advantage of e-commerce, following-up to completion of the development work required, and the preparation of periodic reports on progress of work on a regular basis (Saudi Ministry of Commerce 2001). The committee has prepared a general framework for a plan to apply e-commerce systems in Saudi Arabia. This framework includes the improvement of various factors involved with e-commerce transactions (e.g. IT infrastructure, payment systems, security needs, legislations and regulations, delivery systems etc). The plan also includes the development of e-commerce education and training (Saudi Ministry of Commerce 2001). However, this information was gained from the first publication booklet of Saudi Ministry of Commerce in regards to e-commerce. No other information has been found either on Saudi Ministry of commerce website or its documents to provide further details about this committee and its current role in e-commerce development in the country.

Albadr (2003) mentioned that research has been conducted to examine 42 countries, including Saudi Arabia, in terms of e-commerce current situations and future developments. This examination involves five factors: 1) the prevalence of Internet services including with suitable qualities and costs, 2) the governments trends to adopt and develop e-commerce systems in their countries, 3) electronic security and legislations, 4) attractive business environment (in terms of competition, commercial policies, payment systems, global trade etc), and 5) skilled employees. The results of this study demonstrated that Saudi Arabia, with other seven countries, takes the lowest assessment marks in these five factors and especially the first factor.

Aladwani (2003) conducted a study to report "certain Internet characteristics and e-commerce issues" and come up with challenges that face "the diffusion of the Internet and its application" in the Arab region. Regarding applying e-commerce, he found that there are primary concerns for both business owners/managers and online customers. The Arab business managers take into their consideration "technical obstacles and the attitudes and behaviours of e-commerce consumers". The Internet customers mentioned business's reputation, security, privacy, and legal regulations.

Some studies have been conducted to figure out what makes Saudi Arabia a late adopter in the field of e-commerce. A commercial guide for US companies "Doing Business in Saudi Arabia" (2005) stated some of the barriers that were encountered in the quick implementation of electronic business in the Saudi markets. These inhibitors involved the slowness of Internet services, resistance to social adaptation to a new commercial paradigm, lack of trust of online business, shortage of skilled employees for implementation and maintenance of e-business systems (US 2008 and CITC 2006).

In a qualitative research, Al-Solbi and Mayhew (2005) interviewed 30 representatives of public and private Saudi organizations: "94% of the interviewees said that the most important obstacle to e-commerce is the problem of the lack of individual house addresses". They also found that "there are no clear regulations, legislation, rules and procedures on how to protect the rights of both organizations and their customers" in the e-commerce field.

In 2007, Saudi Communication and Information Technology Commission (CITC) carried out an extensive study to evaluate the current situation of the Internet, and various aspects involving the Internet usage in Saudi Arabia. One of these aspects is e-commerce awareness and activities. As introduced earlier, for the business commercial organizations, they have found that only 9% of Saudi commercial organizations, mostly medium and large companies from the manufacturing sector, are "involved in e-commerce implementation". It is reported that only 4 out of 10 private companies have their own website. This percentage takes on a higher proportion for the larger oil, gas and manufacturing companies. For the users (customer), they have found that 43% of the respondents were aware of e-commerce and only 6% ever bought or sold products online, "mainly airline tickets and hotel bookings" (CITC 2007).

Alfuraih (2008) said: "Universally, e-commerce has three main 'pillars': communication and Internet, payment, and delivery services". Looking at delivery services in Saudi Arabia, the Saudi postal services remain inefficient (Alfuraih 2008). From the establishment of Saudi Post until 2005, individuals had no uniquely identifying home addresses and the mail was not delivered to homes and offices (Saudi Post 2008). If individuals want to receive mail, they have to subscribe to have mail boxes in the local post offices (Alfuraih 2008). In 2005, the new project for addressing and delivery to homes and buildings was announced and approved by Saudi Post (Alfuraih 2008 and Saudi Post 2008).

Agamdi (2008) said: delayed implementation of electronic trading in Saudi Arabia was due to challenges of legislation; organizational management; quality of electronic commerce portals; online payments systems; consumer culture; mailing services, and technical aspects.

Alrawi and Sabry (2009) conducted a study based on a literature review to explore "the status of e-commerce in the Gulf region in relation to international e-commerce including barriers, ICT development and the factors that may drive its success". The possible barriers to e-commerce include users' reluctance due to perceived security and trust-related issues; technological aspects related to Internet connections and the cost of building and managing websites; usability and interactivity issues; educational, ICT literacy and cultural issues; English language issues, and low degree of use of available technologies. However the study shows that there is great potential for e-commerce in this part of the world which requires effort to develop. The authors came up with a list of factors that may drive e-commerce growth in the Gulf countries: "awareness and education levels, users' confidence in online transactions, trust and security issues, usability and interactivity of websites, change management, IT skills development and language fluency, ICT infrastructure, law awareness for e-commerce, assurance and encouragement from government and online companies, endorsement of banks of online payment systems, industry standards and competitive advantage and the establishment of an Arab online business model".

## 3. E-COMMERCE AND THE DIFFUSION OF INNOVATION MODEL

A framework is required to give a way of ordering and looking at aspects of diffusion of innovation as it applies to e-commerce. The Diffusion of Innovation (DOI) model is used as framework to guide this study because of the following reasons. (1) Rogers (2003) defined an innovation as an idea, practice or object that is perceived as new by an individual or other unit of adoption. As this study involves diffusion of the adoption of e-retail systems which are new technology (innovation) for Saudi society, DOI should be considered. (2) Using DOI in this study is believed to be the most relevant, "as a variety of diffusion studies had shown that they consistently influence adoption" (Al-Gahtani 2003). DOI model has been widely believed to 'best' explain such adoption of innovation (Pease & Rows 2005). It has been argued that the diffusion of innovation theory is more relevant to the study of e-commerce due to the technical components of e-commerce (Chong & Bauer 2000; Ling 2001 as cited in Sparling & Caster-Steel 2007). (3) The DOI model helps to look at different aspects (e.g. attributes of innovations, type of innovation-decision, communication channel, nature of the social system, and extent of change agent's promotion efforts) that influence the decision-making to adopt (Rogers 2003). (4) "Diffusion of innovation (DOI) theory provides a solid foundation for developing conceptual models that assess the impact of new information technology on users, over time" (Alkhateeb, Khanfar & Loudon 2010). (5) DOI "provides a framework for analysis of the diffusion of innovations at a complex system level, taking into account the differences in users, rate of adoption, types of information and decisions, and communication channels, while simultaneously facilitating identification of highly specific attributes of innovation that affect diffusion" (Lee 2004 as cited in Alkhateeb, Khanfar & Loudon 2010).

DOI provides an inclusive view of the process of innovation decision making (Rogers 2003). It undertakes to explain how an idea, practice, or object that is perceived as new by an individual or other unit of adoption is spread (Eastin 2002). According to Rogers (2003, p 5), diffusion is "the process during which an innovation is communicated through certain channels over time among members of a social system". With this explanation, Rogers set four elements for the diffusion of innovation. These elements are (1) the innovation, (2) communication channels, (3) time, and (4) social systems (Rogers 2003, p 11). An innovation is "an idea, practice, or object that is perceived as new by an individual or other unit of adoption" (Rogers 2003, p 12). Communication is "the process by which participants create and share information with one another in order to reach a mutual understanding" (Rogers 2003, p18). "The inclusion of time as variable in diffusion research is one of its strengths". The time dimension is "involved in diffusion in (1) the innovation-decision process by which an individual passes from first knowledge of an innovation through its adoption or rejection, (2) the innovativeness of an individual or other unit of adoption (that is, the relative earliness/lateness with which an innovation is adopted) compared with other members of a system, and (3) an innovation's rate of adoption in a system, usually measured as the number of members of the system who adopt the innovation in a given time period" (Rogers 2003, p20). A social system is defined "as a set of interrelated units that are engaged in joint problem solving to accomplish a common goal. The members or unit of a social system may be individual, informal groups, organizations, and/or subsystems." (Rogers 2003, p23)

Rogers (2003) identified five attributes determining the innovation's rate of adoption. He highly recommended that each diffusion research should develop the measures of the five perceived attributes (Rogers 2003, p 256). These five variables are (1) perceived attributes of innovations (Relative advantage, Compatibility, Complexity, Trialability, and Observability), (2) type of innovation-decision (optional, collective, authority), (3) communication channel diffusing the innovation at various states in the innovation-decision process (mass media, interpersonal), (4) nature of the social system (norms, degree of network interconnectedness, etc), and (5) extent of change agent's promotion efforts (Rogers 2003, p 222).

However, there is a claim that e-commerce has unique features distinguished it from other types of innovations (Chan & Swatman 1999; Chong & Bauer 2000; Sparling & Caster-Steel 2007). As a consequence, they proposed different models of e-commerce adoption including some variables from DOI. However, proposing new models of e-commerce adoption and not using DOI alone can be argued that they consider only one type of variable that determines an innovation's rate of adoption; and that is why their argument is weak. They looked at part of the whole picture and ignored others. Rogers (2003, p 222) argues that "the five types of variable that determine an innovation's rate of adoption has not received equal attention from diffusion scholars. The five perceived attributes of innovations have been most extensively investigated and have been found to explain about half of the variance in innovations' rates of adoption".

## 4. RESEARCH METHODOLOGY

This study initially involves exploratory research using a qualitative approach. By adopting a qualitative approach this study is able to gain an in-depth understanding of the e-retail phenomena in KSA. A semi-structured interview was designed based on Diffusion of Innovation theory (DOI).

Interviews were conducted with 16 retailers' decision makers (including owners, headquarter managers, marketing managers, and IT managers) and 16 potential online retailers' customers in Saudi Arabia. The sample of retailers decision makers were selected to cover large, medium and small companies and also to cover different types of retail businesses (telecommunications, computers, sports, supermarkets, restaurants, printing services, Internet services, electrical and electronic products, beauty and body cares, books, watches and clocks, and chocolate and biscuit manufacturing). The samples of potential online retailers' customers conducted with 8 males and 8 females and covered the age group from 16 years to 45 years. Interview questions, answers and discussions were all in Arabic, except for two interviews were in English, and the researcher translated the transcripts into English.

## 5. KEY DRIVERS TO ONLINE RETAILING GROWTH IN SAUDI ARABIA

Developing e-retailing systems growth requires a lot of effort to create an enabling environment in a country where selling and buying online seems to be an innovation. The participants raise many challenges that may inhibit the growth of online retailing systems in Saudi Arabia. It is apparent that retailers in Saudi Arabia are disinclined to adopt e-retailing systems at the moment. At the same time, customers too show low interest in buying online due to the difficulties that they encounter. By providing solutions that address inhibitory factors it will be possible to create a positive online business climate. Through this exploratory research there also arose several enabling factors that would promote online retailing systems growth identified by both retailers and customers in Saudi Arabia.

Seven key drivers are shared between retailers and customers that can help for online retailing growth. They both share the same ideas and suggestions to improve the current situation. They include government support, trustworthy and secure online payment options, educational programs, high speed and low cost Internet connection, owning unique house addresses, the security of having Bricks to support the Clicks and competitive prices.

## 5.1 Government Support

It is not surprising for Saudis to seek government's help. This is because Saudi government plays a central role and people have a tendency to trust what comes through the government. The role that the government can responsibly play in facilitating e-retail is based around three key functions: facilitation, supervision, and control.

It seems that the government role is missing. A phone call was made on 13 Dec 2010 to Saudi Ministry of Commerce to gain more information about e-commerce committee that was established in 2001 as discussed in earlier section. A ministry representative replied "this committee does no longer exist under Ministry of Commerce. Since 2006, the responsibility of e-commerce has transferred to the Ministry of Communications and Information Technology". In 2006, Saudi Ministry of Communications and Information Technology produced a national plan for ICT in Saudi Arabia entitled "The National Communications and Information Technology Plan". The vision statement of this plan is "the transformation into an information society and digital economy so as to increase productivity and provide communications and IT services for all sectors of society in all parts of the country and build a solid information industry that becomes a major source of income" (Saudi Ministry of Communication and Information Technology 2006, p 4). They formulated seven general goals to realize this vision. These seven goals further detailed into 26 specified objectives, 62 implementation policies, and 98 projects (Saudi Ministry of Communication and Information Technology 2006 and 2009). However, this plan does not include specific details that involve e-commerce. It is mainly involved with (public sector) e-government and e-learning and what needs to be done to support its activities. No details provided to show the current situation of e-commerce in Saudi Arabia and what need to be done in the future to facilitate and support its activities.

So, online retailing's rules, regulations, and legislation should be set up by government organization to help protect all involved parties and in return help to protect trust. The current role of government is development so it needs to take a more active and proactive role.

## 5.2 Owning House Addresses

Since the establishment of Saudi Post until 2005, individuals had no addresses and the mail was not delivered to homes and offices (Saudi Post 2008). If individuals want to receive mail, they have to subscribe to have mail boxes in the post offices (Alfuraih 2008). In 2005, the new project for addressing and delivery to homes and buildings was announced and approved by Saudi Post (Alfuraih 2008 and Saudi Post 2008). 'Wasel' is a mail service that enables the customer to receive all their mail at their residence, delivered to their mail box free of charge. Residents in Saudi Arabia have to telephone Saudi Post, visit a Saudi post office, or apply online to apply for getting a mailbox with home physical address. There is also another service called "Wasel Special". This service is for a fee and includes the mail delivery service and six other services: sending mail from a home mailbox, delivery with e-stamp, temporary safe keeping, temporary forwarding, P.O.Box transfer and e-mail notification. However, this service is not yet covering all cities in Saudi Arabia, and currently only in the main cities (Saudi Post 2010). The number of subscribers in the 'Wasel' service reached more than half a million (Alriyadh 2010). This number represents almost 2% of population who own individual houses mailbox.

This service is still new in the Saudi environment and will take time to be recognized and adopted by most of the population; however, Saudi Arabia as compared to the developed world is late providing individual addresses. While the service of providing individual houses mailboxes is now there, the small percentage of population currently owning a mailbox signals the need for more effort to investigate what inhibits citizens' movements towards owning a home mailbox. Problems might be related to the lack of awareness issues of the importance owning a mailbox, or might be related to an issue that the citizens do not trust receiving their mails through this new service. Awareness is another issue that may be raised here. Among the participants in this research, none knew that there was a service delivering mail to individual houses, or knew that that it was possible to obtain direct addresses for their houses with numbers and house names. In all cases, while the service is there, more efforts are needed to motivate the citizens owning house mailboxes and figure out the problems that they face.

## 5.3 Providing Trustworthy and Secure Online Payment Options

Buying online using a credit card is the common option in online shopping; however, getting a credit card from Saudi banks is restricted to those that have a specified monthly income deposited into a customer bank account. Also there are people that have concerns regarding security of using a credit card to buy items online. A third concern that some customers raise involves credit cards' administration annual fees which may be considered as interest or "reba" which is forbidden under Islamic law.

So, providing more options (e.g. debit cards, electronic money like PayPal etc) for online payments motivates people to choose the method that they are most satisfied with to buy online. In addition to the satisfaction, offering more options helps to remove concerns regarding security issues of online payment. By making it available for customers to choose the online payment method that they prefer, of course this offer helps them to select not only the method that they feel happy with but also the method that they feel most secure to use. People are different in terms of the payment methods that they feel satisfied with and feel secure to use. Also, there should be an Arabic interface for any payment portal to help those who have sensitivity to or difficulty understanding the process in English.

Currently in Saudi Arabia there is an e-payment system called SADAD. SADAD is a national electronic bill presentment and payment service provider for the Kingdom of Saudi Arabia. The core mandate for SADAD is to facilitate and streamline bill payment transactions of end consumers through all channels of the Kingdom's Banks. SADAD was launched on October 3rd, 2004" (SADAD 2004). However, this system is seen as an expensive solution to be adopted by small and medium retailers. The cost to adopt this system should be reduced to help online retailers to use it as an online payment system. It is currently more commonly used in e-government activities and this will enhance its perceived security which will help to build trust for online retailers.

## 5.4 High Speed and Low Cost Internet Connection

In the first half of 2010, the number of Saudi Internet users reached 11 million which represents 40% of the population. Broadband subscriptions reached 3.20 million at the same period which represents 11.9% of population (Saudi Ministry of Communication and Information Technology 2010). This is a good indicator of the growth; however, more efforts have to be made in order to make sure that the broadband services are available in more cases and more places. Cities other than main cities in Saudi Arabia are currently not well served with broadband services; this is what participants who live in cities other than main cities complain about. Internet prices also should be at low cost making it available and providing incentive for a wide range of population to access the Internet.

## 5.5 Educational Programs

Online shopping is new to Saudi environment. Educational programs are important in terms of helping remove the exaggerated fear of online purchases, showing the benefit of e-commerce, and clarifying the ways that it can be used to buy and sell online. When people become aware of the online shopping processes, it helps motivate them to buy online. Education is essential in this field and it will necessarily take a while to influence and adopt the culture of e-commerce. Similarly, once the businessmen or those who run the businesses in KSA are confident that people have become keen to go online and to visit commerce sites then they will be more than happy to create their business an e-commerce channel.

Educational programs can be formal or informal and can be delivered in many ways. School is just one way. There are online forums, community education programs, TV programs, interest groups, college programs, etc. All of which can be effective at the right time. Education programs are useful for potential shoppers but they are also a requirement for business people to be able to understand a new form of business and to be able to compete on what becomes a non-local marketplace.

## 5.6 The Trust of having Bricks to Support the Clicks

"Bricks-and-Clicks" is a type of e-retailing system where the online distribution channel for a company complements successful and popular physical stores (Laudon and Traver, 2007, p 69). This type of business seems to be the most preferable type for most of the participants. This type of business reduces the issues of trust as customers know that there is a physical shop they can visit when there is something wrong. It helps to remove the concern of product quality and increases the chance for online purchasing if customers already know the products and their qualities that they are used to buying. Well-known retailers and branded name products' companies gain advantage from this model and their decisions to provide an online sales option motivates more customers to buy online.

## 5.7 Competitive Prices

One of the most important advantages of e-commerce is selling at low cost compared to normal shopping. Offering competitive price is the most attractive of several factors related to buying online. So, online retailers have to think carefully about efficient business processes and business models that support the reduction of online prices compared to normal retail shopping and support shipment fees at a reasonable rate.

## 6. CONCLUSION AND FUTURE WORK

This paper has investigated the issues that positively influence online retailing growth in Saudi Arabia. It comes up with seven key factors driving towards online retailing growth in Saudi Arabia. These factors include government support, providing trustworthy and secure online payments options, motivation to own individual houses mailboxes, providing high speed Internet connection at low cost, providing educational programs, the success of bricks-and-clicks model, and competitive prices. Policy makers and developers may benefit from attention to these factors to facilitate online retailing growth in KSA. However, this paper represents initial findings from an ongoing study that is still in progress. We will in due course be able to develop a more comprehensive formative model in order to contribute to e-commerce development in Saudi Arabia.

## REFERENCES

Agamdi, A 2008, 'e-Commerce Implementation Challenges and Success Factors in the Kingdom of Saudi Arabia', paper presented to 19th National Computer Conference: the digital economy and ICT industry, Riyadh, 1-5 Nov.

Aladwani, AM 2003, 'Key Internet characteristics and e-commerce issues in Arab countries', *Information Technology & People*, vol. 16, no. 1, pp. 9-20.

Albadr, BH 2003, 'E-commerce', Science and Technology, no. 65, pp. 14-9.

Alfuraih, S 2008, 'E-commerce and E-commerce Fraud in Saudi Arabia: A Case Study', in *2nd International Conference on Information Security and Assurance* Busan, Korea, pp. 176-80.

Al-Gahtani, S 2003, 'Computer technology adoption in Saudi Arabia: correlates of perceived innovation attributes', *Information Technology for Development*, vol. 10, no. 1, pp. 57-69.

Alkadi, IA 2008, *Explore the Future of Telecommunications and Information Technology in KSA*, Communications and Information Technology Commission, Riyadh.

Alkhateeb, F, Khanfar, N & Loudon, D 2010, 'Physicians' Adoption of Pharmaceutical E-Detailing: Application of Rogers' Innovation-Diffusion Model', *Services Marketing Quarterly*, vol. 31, no. 1, pp. 116-32.

Al-Otaibi, MB & Al-Zahrani, RM 2003, E-commerce Adoption in Saudi Arabia: An Evaluation of Commercial Organizations' Web Sites, King Saud University, Riyadh, Unpublished Article.

Alrawi, KW & Sabry, KA 2009, 'E-commerce evolution: a Gulf region review', *Int. J. Business Information Systems*, vol. 4, no. 5, pp. 509-26.

Alriyadh 2010, *No E-government without the application of mailing Addressing System* Saudi Post, viewed 14 Dec 2010, <http://www.sp.com.sa/arabic/news/pages/newsdetails.aspx?ItemID=419>

Al-Solbi, A & Mayhew, PJ 2005, 'Measuring E-Readiness Assessment in Saudi Organisations Preliminary Results From A Survey Study', in I Kushchu & MH Kuscu (eds), *From e-government to m-government*, Mobile Government Consortium International LLC, Brighton, UK, pp. 467-75.

Bazeley, P 2010, *Qualitative Data Analysis with NVivo*, Sage Publication Ltd, London

Boeije, H 2010, *Analysis in Qualitative Research*, Sage Publications Ltd, London

Central Inelegance Agency (CIA) 2009, *The World Fact Book: Saudi Arabia*, viewed 20 Oct 2010, <https://www.cia.gov/library/publications/the-world-factbook/geos/sa.html>

Chan, C & Swatman, P 1999, 'E-Commerce Implementation in Australia: a Case Study Approach', in *The 3rd Collaborative Electronic Commerce Technology and Research* Wellington, New Zealand, pp. 1-13.

Chong, S & Bauer, C 2000, 'A model of factor influences on Electronic Commerce adoption and diffusion in small-and medium-sized enterprises', in *The 18th Pacic Asia Conference on Information Systems (PACIS)*, Shanghai, China.

CITC (Communications and Information Technology Commission) 2007, *Internet Usage Study in the Kingdom of Saudi Arabia* Communications and Information Technology Commission, Riyadh

Eastin, M 2002, 'Diffusion of e-commerce: an analysis of the adoption of four e-commerce activities', *Telematics and informatics*, vol. 19, no. 3, pp. 251-67.

Information Centre - Saudi Ministry of Commerce 2006, 'E-commerce in the Kingdom of Saudi Arabia ', paper presented to Arab Organization for Industrial Development Conference Tunisia, 19-21 April.

Johnson, RB & Onwuegbuzie, AJ 2004, 'Mixed Methods Research: A Research Paradigm Whose Time Has Come', *Educational Researcher*, vol. 33, no. 7, pp. 14-26.

KACST & Saudi Ministry of Economy and Planning 2009, *Strategic Priorities for Information Technology Program*, King Abdulaziz City for Science and Technology, Riyadh

King, N & Horrocks, C 2010, *Interviews in Qualitative Research*, SAGE Publications Ltd, London.

Laudon, KC & Traver, CG 2007, *E-commerce: business, technology, society*, 3 edn, Pearson Prentice Hall, New Jersey.

Neuman, WL 2006, *Social Research Methods: Qualitative and Quantitative Approaches*, 6th edn, Pearson Education, Boston, MA.

Pease, W & Rowe, M 2005, 'Diffusion of Innovation - The Adoption of Electronic Commerce by Small and Medium Enterprises (SMES)- A Comparative Analysis', *Australasian Journal of Information Systems*, vol. 13, no. 1, pp. 287-94.

Rogers, EM 2003, *Diffusion of Innovations*, Fifth edn, Simon & Schuster, New York

SADAD 2004, *About SADAD Payment System*, viewed 10 Oct 2010, <http://www.sadad.com/English/SADAD+SERVICES/AboutSADAD>

Sait, SM, Al-Tawil, KM & Hussain, SA 2004, 'E-commerce in Saudi Arabia: Adoption and Perspectives', *Australasian Journal of Information Systems*, vol. 12, no. 1, pp. 54-74.

Saudi Ministry of Commerce 2001, *E-commerce in the kingdom: Breakthrough for the future, Saudi Ministry of Commerce*, Riyadh, (Arabic source).

Saudi Ministry of Communication and Information Technology 2006, *The National Communications and Information Technology Plan*, Ministry of Communication and Information Technology, Riyadh.

Saudi Ministry of Communication and Information Technology 2009, *The Annual Report for the National Communications and Information Technology Plan*, Ministry of Communication and Information Technology, Riyadh.

Saudi Ministry of Communication and Information Technology 2010, *ICT Indicators in the Kingdom of Saudi Arabia (1st Half- 2010)*, Ministry of Communication and Information Technology, viewed 15 Dec 2010, <http://www.mcit.gov.sa/english/Development/SectorIndices/>

Saudi Post 2008, *Saudi Post: Establishment and Development*, Saudi Post, viewed 21 Nov 2009, <http://www.sp.com.sa/Arabic/SaudiPost/aboutus/Pages/establishmentanddevelopment.aspx>

Saudi Post 2010, *Production and Services*, Saudi Post viewed 14 Dec 2010, <http://www.sp.com.sa/English/SaudiPost/ProductsServices/Pages/Wasel1.aspx>

Sparling, L, Cater-Steel, A & Toleman, M 2007, 'SME adoption of e-Commerce in the Central Okanagan region of Canada', in 18th Australasian Conference on Information Systems, Toowoomba, pp. 1046-59.

U.S. Department of Commerce 2008, *Doing Business In Saudi Arabia: A Country Commercial Guide for U.S. Companies*, U.S. & Foreign Commercial Service and U.S. Department of State.